\documentclass[%
 aip,
 amsmath,amssymb,
 reprint,%
]{revtex4-1}
\newcommand{\sbs}{Sb${_2}$S${_3}$ }
\newcommand{\sbse}{Sb${_2}$Se${_3}$ }
\newcommand{\gst}{Ge${_2}$Sb${_2}$Te$_5$ }
\newcommand{\gsst}{Ge${_2}$Sb${_2}$SeTe$_4$ }

\usepackage{graphicx}
\usepackage{dcolumn}
\usepackage{bm}
\usepackage[utf8]{inputenc}
\usepackage[T1]{fontenc}
\usepackage{mathptmx}
\usepackage{etoolbox}
\usepackage{siunitx}
\usepackage{color,soul}
\makeatletter
\def\@email#1#2{%
 \endgroup
 \patchcmd{\titleblock@produce}
  {\frontmatter@RRAPformat}
  {\frontmatter@RRAPformat{\produce@RRAP{*#1\href{mailto:#2}{#2}}}\frontmatter@RRAPformat}
  {}{}
}%
\makeatother
\begin{document}

\preprint{AIP/123-QED}

\title{Low-power 7-bit hybrid volatile/ nonvolatile tuning of ring resonators}

\author{Jayita Dutta}
\altaffiliation{Electrical and Computer Engineering (ECE) Dpt.,University of Washington}
\author{Rui Chen}
\altaffiliation{Electrical and Computer Engineering (ECE) Dpt.,University of Washington}
\author{Virat Tara}
\altaffiliation{Electrical and Computer Engineering (ECE) Dpt.,University of Washington}
\author{Arka Majumdar}
\email{jayitad@uw.edu , arka@uw.edu}
 \altaffiliation{Electrical and Computer Engineering (ECE) Dpt.,University of Washington}
\altaffiliation{Department of Physics,University of Washington}
\date{\today}

\begin{abstract}
Programmable photonic integrated circuits are expected to play an increasingly important role to enable high-bandwidth optical interconnects, and large-scale in-memory computing as needed to support the  rise of artificial intelligence and machine learning technology. To that end, chalcogenide-based non-volatile phase-change materials (PCMs) present a promising solution due to zero static power. However, high switching voltage and small number of operating levels present serious roadblocks to widespread adoption of PCM-programmble units. Here, we demonstrate electrically programmable wide bandgap Sb$_2$S$_3$-clad silicon ring resonator using silicon microheater at CMOS compatible voltage of < 3V. Our device shows low switching energy of 35.33 nJ (0.48 mJ) for amorphization (crystallization) and reversible phase transitions with high endurance (> 2000 switching events) near 1550 nm. Combining volatile thermo-optic effect with non-volatile PCMs, we demonstrate 7-bit (127 levels) operation with excellent repeatability and reduced power consumption. Our demonstration of low-voltage and low-energy operation, combined with the hybrid volatile-nonvolatile approach, marks a significant step towards integrating PCM-based programmable units in large-scale optical interconnects.
\end{abstract}

\maketitle
\section{\label{sec:level1}INTRODUCTION}

To support the rising demands of compute for artificial intelligence (AI), there is a clear need to transform both computing and communication systems\cite{zhu2023intelligent,christensen20222022,gill2022ai,mayer2020scalable,kelechi2020artificial,markov2014limits}. On one hand, a large amount of data needs to move between electronic computers and servers. Optical interconnects present a promising solution thanks to their much lower latency and energy consumption compared to the traditional copper-based interconnects \cite{thraskias2018survey,tsiokos20162,tsiokos2017optical,haffner2015all}.  Especially, optical circuit switching systems are already being deployed in data-centers \cite{urata2022mission,liu2021bridging}, as recently announced by Google. These switches are slow (e.g., Google is deploying micro electro-mechanical switches), but need to provide very large change in optical properties, with ultra-low loss. On the other hand, the growing demand for vector-matrix multiplication \cite{amirsoleimani2020memory,woods2017approximate,maslej2023artificial,zhou2022photonic} poses a serious challenge for traditional computing architectures, including graphical processing units. This has led to exploration of optical vector-matrix multiplication, which requires in-memory computing \cite{hattori2021neural,athale1982optical,ocampo2024new}.  Both  applications, optical circuit switching and vector-matrix multiplication, require ultra-compact, low-loss programmable photonic units \cite{chen2024low}, with high endurance, compatibility with electronic integrated circuit (i.e., the actuation electrical power can be provided by on-chip circuit) and high bit-precision \cite{zhu2020photonic,chakraborty2019photonic,lian2022photonic}. 

While large-scale photonic integrated circuits (PICs) have now become available\cite{shen2017deep,capmany2016programmable,harris2017quantum,zhu2021integrated,prabhathan2023roadmap,moody20222022,schulz2024roadmap,bogaerts2020programmable,haselman2009future}, thanks to sophisticated fabrication techniques, a truly programmable photonic circuits with the aforementioned properties remain missing \cite{geler2022ferroelectric,gao2021intermediate,moitra2023programmable,li2019fast}. We emphasize that, most of the existing reconfiguration methods in integrated photonics are volatile (thermo-optic, free-carrier dispersion and Pockels effect\cite{perez2017multipurpose,zhang2020photonic,reed2010silicon,wang2018integrated,melikyan2014high,prencipe2023electro}), and are essentially optimized for high-speed switching. However, both optical circuit switching and vector-matrix multiplication, will benefit from programmable photonic units, which are changed infrequently, but have zero static power \cite{schrauwen2008trimming,feldmann2021parallel,wu2021programmable,wan2020large,zhang2021myths}. Chalcogenide phase-change materials (PCMs) offer a promising solution to more compact device size and zero static power owing to their strong optical index modulation ($\Delta n \sim O(1)$) and zero static power consumption \cite{wuttig2017phase,abdollahramezani2020tunable,fang2021non, chen2023non}.  

PCMs possess two stable, reversibly switchable micro-structural phases namely amorphous phase (a-phase) and crystalline phase (c-phase), with significantly different optical refractive indices ($\Delta n$ $\sim$ 1). Due to their non-volatile phase transition under the ambient environment, PCMs do not require static power to maintain their state once switched. The large refractive index change ($\Delta n$) and non-volatility allow for the creation of compact reconfigurable devices ($\sim$ \SI{100}{\micro\metre}) \cite{chen2022opportunities,rios2022ultra} with zero static energy consumption \cite{zheng2018gst,xu2019low,fang2021non,chen2022broadband,fang2022ultra,zheng2020nonvolatile}. Furthermore, PCMs are suitable for large-scale integration as they can be easily deposited by sputtering \cite{zheng2018gst,fang2021non,chen2022broadband,zheng2020nonvolatile,rios2015integrated,erickson2022designing,chen2024deterministic,chen2024toward} or thermal evaporation \cite{rios2022ultra} onto various integrated photonics material platforms, including silicon and silicon nitride. While traditional PCMs like \gst (GST) and GeTe show strong optical absorption, emerging wide-bandgap PCMs, such as \gsst (GSST) \cite{zhang2019broadband}, \sbs \cite{delaney2020new,vcernovskova2014thermal}, and \sbse\cite{dong2019wide,nasr2016first,gutierrez2022interlaboratory,wei2023electrically}, offer new opportunities to reduce this absorption loss. These wide bandgap PCMs have garnered significant interest due to potential for both large-area endurable switching and repeatable multilevel operation \cite{rios2022ultra,fang2022ultra,shalaginov2021reconfigurable,lu2021reversible,moitra2023programmable,zhang2018broadband}. However, even though \sbse is also a promising PCM, Selenium is highly toxic in nature \cite{spallholz1994nature}. A recent paper showed tuning of \sbse between its amorphous and crystalline states with PIN heater \cite{wei2023electrically}, which requires high switching voltage (8.2 V for amorphization and 3.3 V for crystallization), incompatible with the standard driving voltage of Complementary-Metal-Oxide-Semiconductor (CMOS) electronic integrated circuits. Although the work showed both PCM and thermal tuning of the micro-ring resonators separately, the collective tuning was not demonstrated.  

Notably, \sbs has the widest bandgap among these emerging PCMs, providing transparency down to $\sim$ 600 nm \cite{dong2019wide} in the amorphous phase. A previous experiment with \sbs, showed low-loss (<1 dB), endurable (>1000 cycles) and 5-bit operation by tuning PCM \sbs \cite{chen2023non} and also its potential in high-volume manufacturing in 300-mm wafers\cite{chen2024deterministic}. However, the switching energy of the devices  remained high, mainly limited by the sub-optimal design of our p++-i-n++ (PIN) doped silicon microheaters \cite{zhang2023nonvolatile}. Moreover, due to inherent stochastic switching behavior of the PCMs, the number of operation levels are limited \cite{tuma2016stochastic}. 

In this work, we show both low-energy switching of PCM \sbs and a large number of intermediate levels at optical telecommunication wavelengths (C-band, $\sim$ 1500 nm –1560 nm). The low switching energy is achieved by optimizing our PIN microheater geometry. Both numerical and experimental results show that the switching energy monotonically reduces with the decrease of the intrinsic region width of the PIN microheater, albeit at the cost of a lower quality ($Q$) factor. An intrinsic region width of \SI{0.9}{\micro\metre} is chosen to achieve a good tradeoff, showing low switching voltage less than 3 V ($\sim$ 25.7 mA current and 500 ns pulse duration, 35.33 nJ) and 1.6 V ($\sim$ 15 mA current and 20 ms pulse duration, 0.48 mJ energy) for amorphization and crystallization  respectively. This voltage and current can be easily provided by CMOS electronic integrated circuits. We show that such low power operation do not degrade the cyclability by demonstrating >10000 switching events in total on a a single device. Finally, we demonstrate $\sim 7$-bit operation (127 levels)  using hybrid thermo-optic/PCM tuning, combining a coarse $\sim 5$ bit (22 levels) non-volatile tuning by PCM and volatile finer tuning by thermo-optic. The CMOS compatible voltage/energy operation coupled with $\sim 7$-bit operation marks important steps towards PCM-based large-scale PIC systems \cite{fang2023non,prabhathan2023roadmap}. 

\begin{figure*}[h!tbp]
\includegraphics[width=\textwidth]{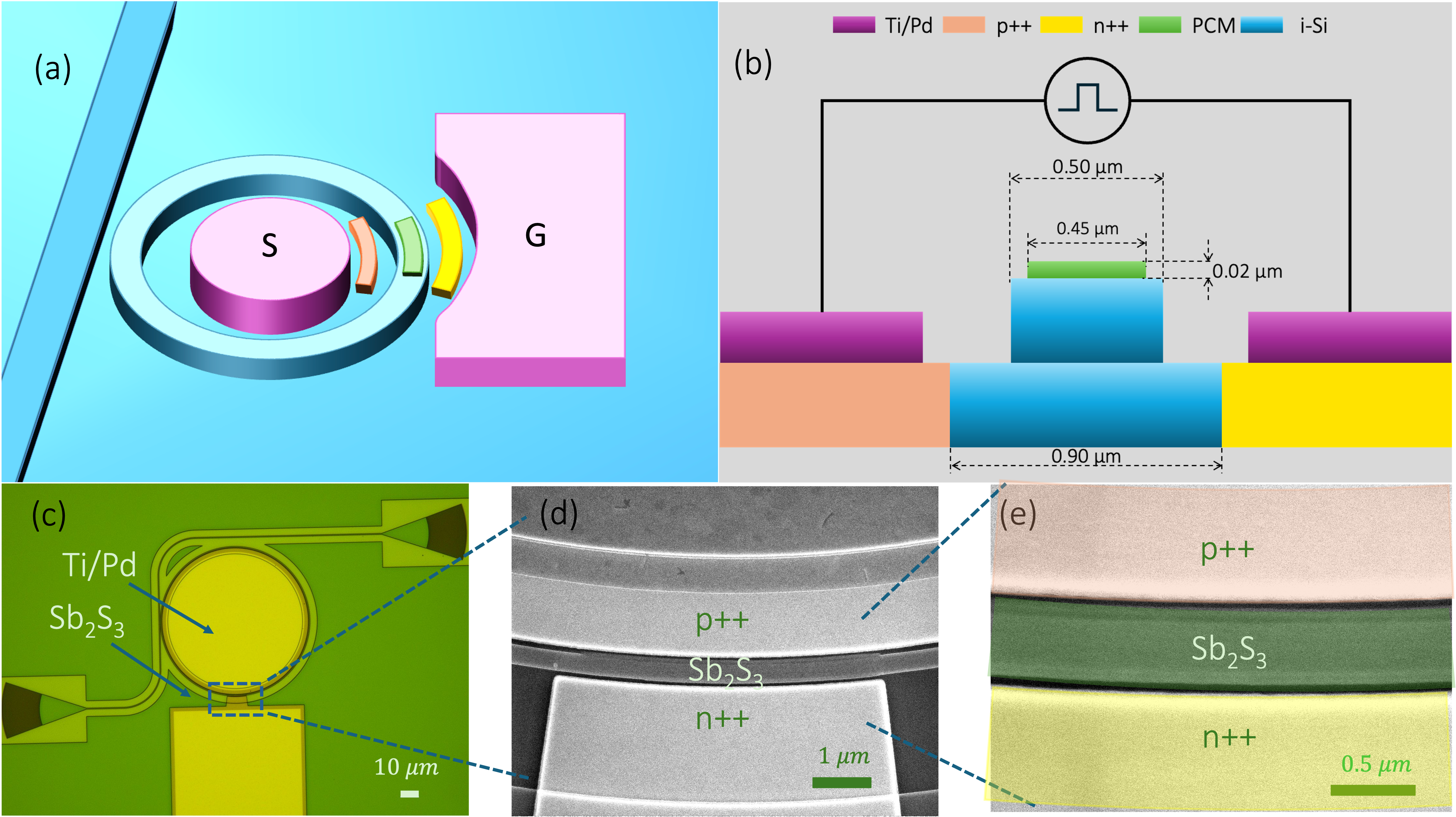}
\caption{\label{fig:1} A high Q ring resonator loaded with \SI{10}{\micro\metre} long 20 nm thick ${Sb_2}S_3$ \textbf{a.} A schematic of the \sbs loaded micro-ring resonator with low operation voltage and energy \textbf{b.} A 2D schematic of the \sbs loaded p++ - i - n++ (PIN) micro-heater design \textbf{c.} Optical microscope image of the micro-ring resonator \textbf{d.} Scanning electron microscope (SEM) image of \sbs loaded PIN micro-heater \textbf{e.} Zoomed SEM view of \sbs loaded PIN micro-heater. The \sbs thin film, i-Si, n++, and p++ doped silicon regions are represented by false colors (green, blue, yellow, and orange respectively)}
\end{figure*}

\section{\label{sec:level1}RESULTS}
We use micro-ring resonators (MRRs) as the platform to demonstrate low-voltage, low-energy operation. Figure \ref{fig:1}a shows the schematic of a silicon micro-ring resonator cladded with 10-$\mu$m-long, 0.45-$\mu$m-width and 20-nm-thick Sb$_2$S$_3$, and the cross-sectional view is on the right. The MRRs were fabricated on a standard silicon-on-insulator (SOI) wafer with a 220 nm silicon layer and a \SI{2}{\micro\metre} buried oxide layer. All the rings have a bus-ring gap of 270 nm to achieve a near-critical coupling condition. The 500 nm wide waveguides were created by partially etching 120 nm of silicon. We then deposited 450 nm wide \sbs stripes onto the SOI chip via sputtering and liftoff. The slightly smaller width compared to the waveguide compensates for the electron beam lithography (EBL) overlay tolerance and error from liftoff. The \sbs films are electrically controlled via on-chip silicon PIN micro-heaters \cite{chen2023non,chen2022broadband,zheng2020nonvolatile}, of which the intrinsic region width were varied from 0.9 to 1.9 $\mu$m to study its relationship with operation voltage, power and Q-factor. The \sbs stripes are encapsulated with 40 nm of thermal Al$_2$O$_3$ grown by atomic layer deposition (ALD) at 150°C. This conformal encapsulation is critical for preventing \sbs from oxidation and thermal reflowing, which is essential for high endurance \cite{chen2023non}. Figures \ref{fig:1}c-d show the optical micrograph and scanning electron microscope (SEM) image of the MRR, respectively. Figure \ref{fig:1}e shows a zoomed SEM image of PIN micro-heater where the doping regions and \sbs are false colored.

\subsection{\label{sec:level2} Optimization of operation voltage and switching energy by varying the intrinsic region width}
\begin{figure*}[h!tbp]
\includegraphics[width=\textwidth]{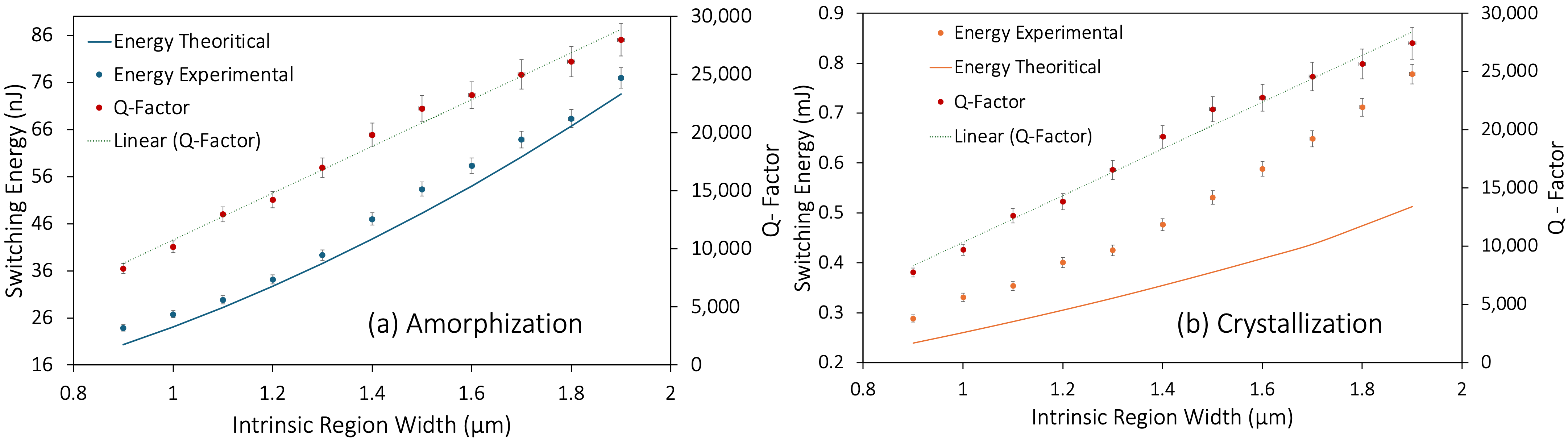}
\caption{\label{fig:2} Optimization of switching energy and Q-factor of \sbs by varying the intrinsic region width of the PIN heater from 0.9  $\mu$m to 1.9 $\mu$m. \textbf{a.} Variation from crystalline to partial amorphous state \textbf{b.} Variation from partial amorphous state to crystalline state. In both Figs. \ref{fig:2} (a-b), green dotted line acts as a guide to the eye to show gradual increase in the Q-factor with the increase in intrinsic region width from 0.9 to 1.9 $\mu$m.}
\end{figure*}

We start with heat transfer simulations in COMSOL Multiphysics, where the intrinsic region width ($w_i$) of the PIN microheater is varied to optimize the switching energy. The COMSOL simulation couples semiconductor and heat transfer modules, and therefore can capture the main physics of our device. The solid lines in Figs. \ref{fig:2}a and \ref{fig:2}b show the simulation results for $w_i$ within 0.9$\mu$m to 1.9 $\mu$m. We clearly see a monotonic reduction of the switching voltage and energy with the decrease of $w_i$. The increase of the amorphization voltage from 3 V at 0.9 $\mu$m to 5V at 1.9 $\mu$m shows the importance of careful designing of the microheater geometry. We attribute this monotonic trend to two factors: (1) the increase in $w_i$ increases the PIN diode resistance and reduces the current flow; (2) the overall heating volume of the PIN microheater increases, thus requiring more energy. Importantly, the simulation shows that at $w_i=0.9$ $\mu$m, pulses with 3 V amplitude (13.53 mA current) and 500 ns duration is already enough to amorphize the material. The \sbs is heated above its melting temperature ($801\pm 18$ K\cite{dong2019wide}), after which it is quenched below the glass transition temperature of 573 K in 200 ns (quench rate $\sim$2 K/ns).
\begin{figure*}[h!tbp]
\includegraphics[width=\textwidth]{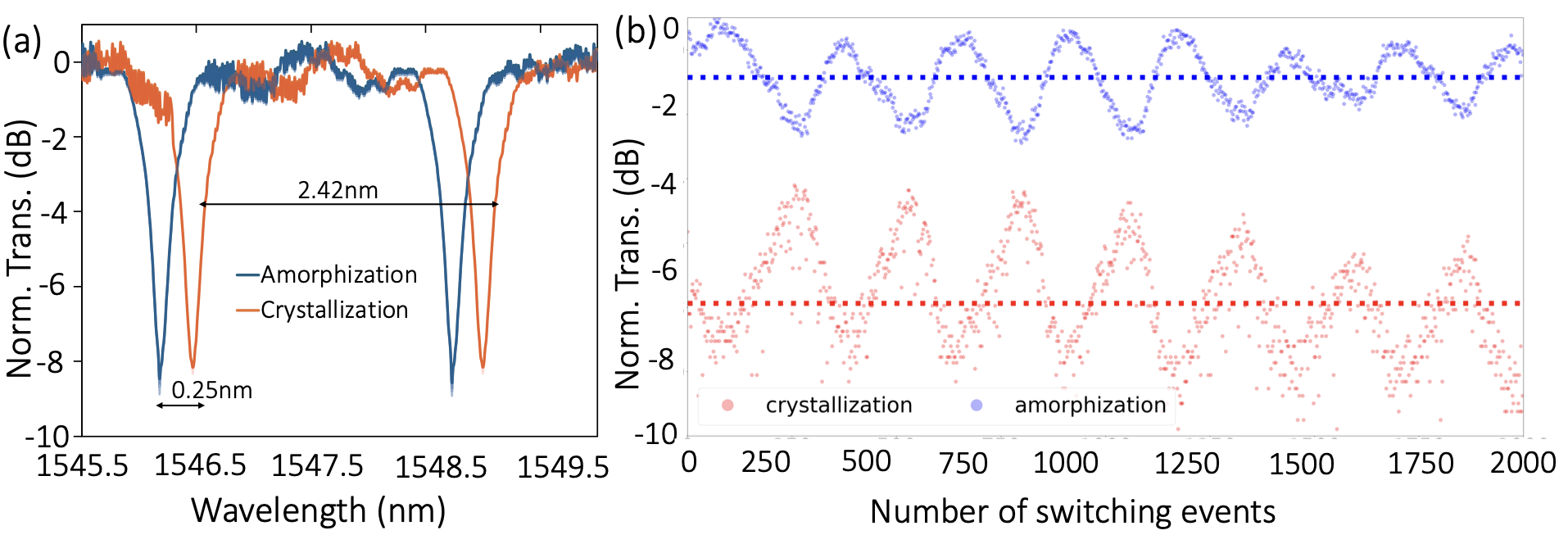}
\caption{\label{fig:4} Low power switching and endurance of ${Sb_2}S_3$ loaded micro-ring resonator with $w_i=0.9$   $\mu$m \textbf{a.} Measured MRR spectra for crystallization (SET) and amorphization (RESET) (SET:  1.6 V, 20-ms-long, 0.48 mJ pulse to change \sbs to the crystalline state; RESET:  2.75 V, 500-ns-long, 35.33 nJ pulse to change \sbs to the amorphous state). The spectra are averaged over five cycles, and the shadowed area shows the standard deviation. Norm. Transmission stands for normalized transmission, normalized to a reference waveguide on the same chip \textbf{b.} Measured endurance for 2000 switching events (1000 crystallization and 1000 amorphization) with the same phase change conditions as mentioned in Fig. \ref{fig:4}a. The blue and orange scatterers represent the normalized transmission when \sbs is in the amorphous and crystalline phases respectively. The observed sinusoidal oscillation pattern in the endurance test is attributed to the self-heating effect of the silicon micro-ring resonator. The average transmission after amorphization and crystallization are indicated by the dotted lines. Norm. Trans. stands for normalized transmission, which is calculated as Transmission (dB) = 10 $\times log_{10}$ (Transmitted Power in mW) and the maximum transmission is normalized to 0 dB. }
\end{figure*}

 Next, we experimentally validate our simulation results (the measurement results are shown by the circles in Fig. \ref{fig:2}). The trend we observed in experiment agrees qualitatively well with our simulation. The SEM images corresponding to different $w_i$ are shown in Supplementary Section 1 (see Supplementary Fig. 1). Experimentally, we have measured the switching conditions for partial amorphization for a fixed resonance wavelength shift of 0.12 nm (i.e, $\Delta \lambda$=0.12 nm) and not the maximum possible $\Delta \lambda$. This choice was made to ensure that the device is not damaged while finding the amorphization and crystallization conditions and all the devices with varying parameters can be measured. However, it is important to note that complete amorphization of \sbs will require higher switching voltage and energy. For p++ - i - n++ heater with $w_i=0.9$   $\mu$m, complete amorphization of \sbs will require higher voltage and energy by 10\% and 47\% respectively as shown in Section B and Supplementary Section 2. The switching energy requirement is calculated as switching voltage multiplied by the estimated current from measured IV curve multiplied by applied pulse duration. Further, as shown in Figs. \ref{fig:2}(a-b) the Q-factor decreases with the decrease in intrinsic region width and constitutes the design trade-off. As shown in Figs. \ref{fig:2}(a-b) the energy requirement is slightly higher experimentally as compared to switching conditions computed for the center of PCM via COMSOL simulation (see Supplementary section 2-3, Fig. 2-3). The observed higher energy requirement can be attributed to four factors: (1) in simulation we have considered simplified 2D geometry to calculate the phase change requirement for the centre of PCM but the actual geometry can be more complex, leading to non-uniform heat distribution and potentially higher energy requirements; (2) the actual kinetics of the phase-change process, including nucleation and growth rates, may be different in the experiment compared to the idealized rates assumed in simulations, requiring more energy to achieve a full phase transition in the experiment; (3) the interfaces between the PCM and surrounding materials can introduce additional thermal resistance or electrical contact resistance in reality as compared to the ones captured in simulation, leading to higher energy requirements in experiments; (4) in reality there can be material inhomogeneity in terms of crystalline structure, or defect density, which can lead to different thermal and electrical properties compared to the idealized conditions used in simulations \cite{laprais2024reversible, aryana2023toward, bal2021nucleation}. The variation in Q-factor and energy requirement with respect to variation in intrinsic region width measured across multiple devices is shown in the form of error bars in the Figs \ref{fig:2}(a-b).

\subsection{\label{sec:level2}Low power actuation and endurance of ${Sb_2}S_3$ clad Si MRR }
We then show more characterization results for the lowest voltage and energy device with $w_i=0.9$ $\mu$m (Figs. \ref{fig:1}c–e).  The resonance spectra from 1545.5 nm to 1550 nm are shown in Fig. \ref{fig:4}a (see supplementary section 4 for 1500 nm to 1560 nm). A resonance shift of $\sim$0.25 nm is obtained (Fig. \ref{fig:4}a) upon switching the \sbs from the a- (blue) to the c-phase (orange). Considering the free spectral range of $\sim$ 2.42 nm, this resonance shift amounts to 0.2$\pi$ optical phase shift per round trip. The observed  condition for complete amorphization (RESET pulse) of \sbs is 500 ns pulse width at 2.75V which corresponds to a current of 25.7 mA and switching energy of 35.33 nJ. The crystallization condition (SET pulse) being 20 ms pulse at 1.6 V corresponding to a current of 15 mA and switching energy of 0.48 mJ. The SET and RESET processes were repeated for 5 cycles and the slight variation is shown as standard deviation among the cycles by the shaded region in Fig. 3a, showing excellent repeatability for the binary representation. The IV characteristics for the p++ - i - n++ microheater is shown in supplementary section 5 (Supplementary Fig. 5). The applied amorphization pulse produces very high temperature which further activated the doped silicon. We observed that the condition became stable after around first ten pulses. It is also noted that the device damage happened much more often during crystallization than amorphization due to the long pulse duration and low heater resistance. Even slight increase of the voltage can lead to significant increase in temperature, which can melt the silicon waveguides (see supplementary section 6 for SEM on damaged devices).

We show the endurance test result of the low-voltage, low-energy device in Fig. 3b, demonstrating 2000 switching events (1000 switching cycles) with a pulsing rate of 10 Hz. We performed the endurance test in Fig. 3b by alternatively switching the device between the crystalline and amorphous state of the PCM. The laser was parked at $\sim$1546.25 nm, which is on resonance with the micro-ring cavity for amorphous \sbs. We observed no performance degradation at the end of 2000 switching events, and the experiment was stopped only because of the long duration of the experiment. We noticed a sinusoidal modulating pattern, which can be attributed to the self-heating effect of the micro-ring resonator. To verify that this modulation is not related to PCM modulation, we repeated the cyclability test at a lower pulsing rate of 1 Hz (see supplementary section 7). The frequency of the modulation remains the same (note the x-axis is in different time scale), indicating this modulation is not related to the electrical pulses nor the PCMs. 

Here we provide a more detailed explanation on the constant oscillation. In a doped silicon micro-ring resonator, light is absorbed due to free-carrier absorption or two photon absorption, producing extra heat. Such self-heating effect changes the effective refractive index of the ring by the thermo-optic effect, causing a shift of the ring resonance cite{novarese2022study,bao2023harnessing}. During the cyclability test, we initially parked the laser on resonance, leading to a strong light field in the ring. Due to the self-heating effect, the resonance is gradually shifted until the laser becomes completely off-resonant when light inside the resonator becomes too weak to heat up the cavity. The resonator then gradually cools itself down to the room temperature, causing the laser on-resonance again and repeating the previous process. Such cyclic phenomenon results in the sinusoidal pattern seen in the endurance test in Fig. 3b. We note that this undesired effect can be weakened by reducing the laser power or by setting the laser slightly off-resonance. In our later measurements (Fig. S10), such an effect was eliminated by employing both practices.

\subsection{7-bit multi-level hybrid TO/PCM tuning }\label{sec:level2}
\begin{figure*}
\includegraphics[width=\textwidth]{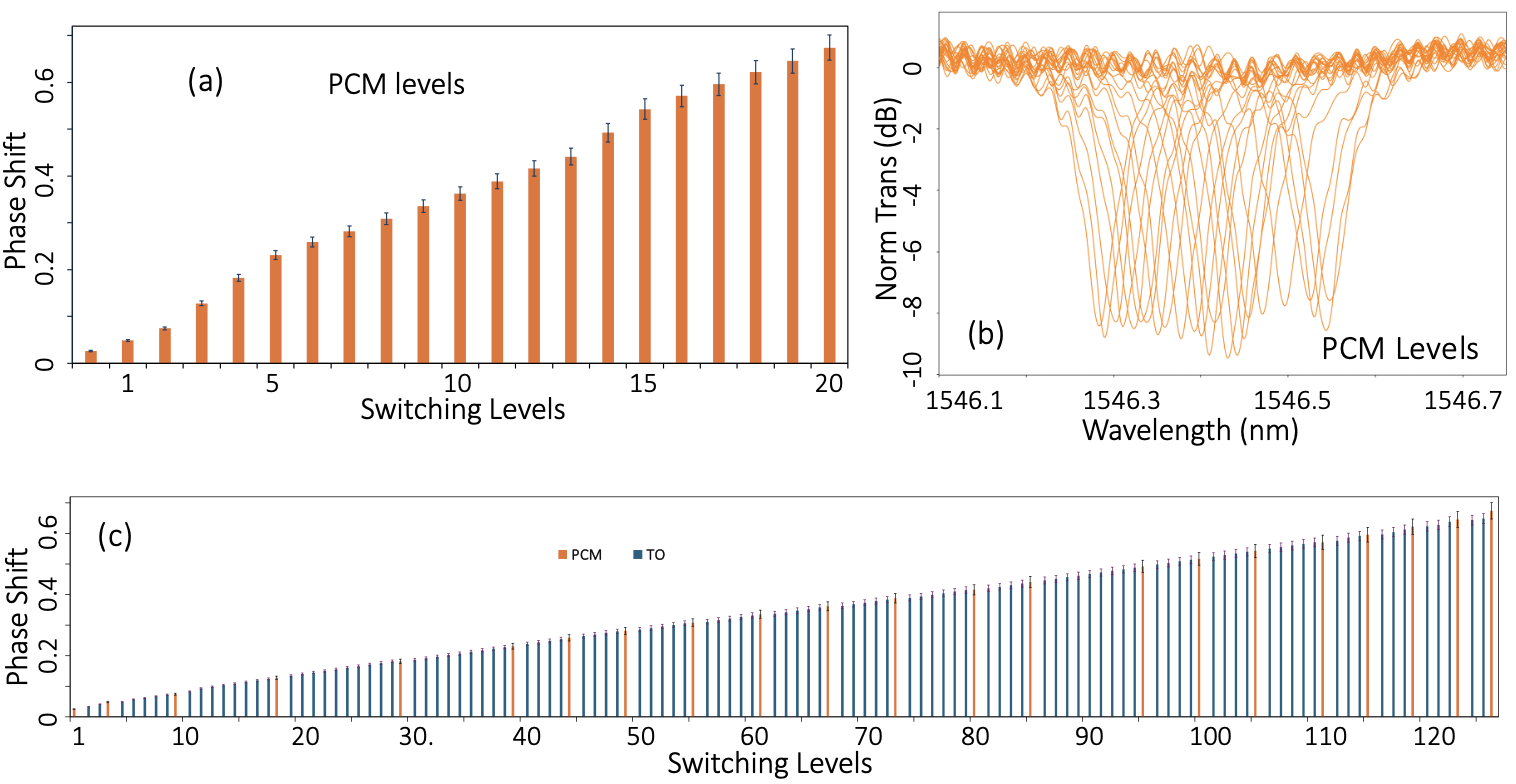}
\caption{\label{fig:3}Seven-bit tuning of micro-ring resonators using hybrid thermo-optic/ PCM tuning.\textbf{a.} Measured phase shifts by multilevel tuning of PCM \sbs over 22 switching levels by applying partial amorphization pulse with voltages ranging from 2.45 to 2.65V, providing resonance shift of $\Delta \lambda$ = 0.01 nm in each step. The partial amorphization pulse changes the ${Sb_2}S_3$ optical phase in a coarse but nonvolatile fashion to achieve a phase shift of 0.02$\pi$ in level 1 and going up to 0.2$\pi$ at level 22 \textbf{b.} The individual transmission spectra due to PCM tuning in the wavelength range of 1546.25 nm to 1546.5 nm \textbf{c.} Measured phase shifts by multilevel tuning of PCM \sbs combined with TO effect to obtain 127 switching levels, thus demonstrating a 7-bit multilevel operation. The PCM tuning is achieved by applying partial amorphization pulse with voltages ranging from 2.45 to 2.65V. In between two PCM tuned levels, precise thermal tuning (5 to 6 TO levels on average) is achieved using the same PIN diode heater by gradually increasing the DC voltage from 0.85V to 1.5V. The orange and blue bars represent the phase shifts obtained due to PCM tuning and TO tuning, respectively.  Norm. Trans. stands for normalized transmission, which  is calculated as - Transmission (dB) = 10 $\times log_{10}$(Transmitted Power in mW) and the maximum transmission is normalized to 0 dB. }
\end{figure*}

The inherent stochasticity of PCM switching limits the achievable number of levels, which can be compensated for by adding thermo-optic fine tuning on top of the PCM coarse tuning. In this configuration, the PIN diode heaters are used for both volatile and non-volatile tuning. With this hybrid tuning approach, we demonstrate 7-bit multilevel operation (up to 127 levels). We started the experiment by applying a 500 ns pulse at 2.45V to partially amorphize the c-\sbs device to demonstrate multiple levels. After one partial amorphization pulse, the resonant wavelength shifts to the left by 0.01 nm (i.e. $\Delta \lambda$ = 0.01 nm) thus corresponding to an intermediate PCM level. The resonance shifts due to PCM tuning enlarges as we increase the amorphization voltage from 2.45 to 2.65V, providing resonance shift of $\Delta \lambda$ = 0.01 nm in each step and 22 operation levels. The partial amorphization pulse changes the \sbs optical phase in a coarse but nonvolatile fashion. Furthermore, we show 350 cycles in total for seven intermediate amorphization levels with only minimal variation in Supplementary Fig. S10. To provide precise thermal tuning, we then applied DC voltage on the same PIN diode heater starting from 0.85 V up to 1.5 V. The resonance wavelength shifts to left at the beginning because the free-carrier dispersion effect dominates the thermo-optic effect. When the applied voltage is larger than 1V, red shifts appeared. This shift in $\Delta \lambda$ is volatile and the resonance shifts back upon removal of the DC voltage. Figure \ref{fig:3}c shows 127 switching levels, with error bars indicating the range from the thermo-optic tuning and PCM tuning over 2 cycles. The thermo-optic levels were recorded by 5 repeated experiments in each cycle to ensure that the resonance shift is from the thermo-optic tuning and not from the PCM tuning. Figure \ref{fig:3}a shows the phase shift for different switching levels via PCM tuning. The individual transmission spectra corresponding to each resonant wavelength shift due to PCM tuning for the first cycle is presented in Fig. \ref{fig:3}b. Another repeatable transmission spectra tuned by PCM is demonstrated in Supplementary Fig. S9. In between two PCM tuned levels, approximately 5 thermo-optic levels are recorded in average by gradually increasing the DC voltage in the range of 0.85 to 1.5V. The phase shift ($\Delta \phi$) is calculated as 

\begin{equation}
\Delta \phi = ({\Delta \lambda}/FSR) \cdot 2 \pi \label{eq:mynum1}
\end{equation}

where, $\Delta\lambda$ corresponds to the resonant wavelength shift for a free spectral range (FSR) of 2.42 nm. The transmission spectra corresponding to the combined effect of thermo-optic and PCM tuning is shown in the Supplementary Section 8[Supplementary Figure 8]. It is important to note that the thermo-optic levels show a much higher loss as corresponding to the PCM levels mainly due to free carrier effect. Further, we would like to note that in the process of doing additional multilevel endurance experiments on the same device, we were able to switch the PCM for > 10K events combining partial and complete amorphization and crystallization cycles.
\section{\label{sec:level1}Discussion}

In the future, the switching voltage and energy can be further reduced by suspending the p-i-n micro-heater, which prevents the generated heat from dissipating to the buried oxide and silicon substrate. This technique was used in thermo-optic phase shifters, exhibiting great efficiency power reduction of > 98\%. Recently, this has been proposed for PCM switching in partially suspended silicon micro ring resonator. \cite{han2023energy}, but no experimental demonstration has yet been reported. We note that one of the main challenges of this approach is engineering the structural stability to avoid deformation or thermal damage during the fabrication and operation. Furthermore, using multiple PCM islands instead of one single PCM patch might further improve the endurance of phase-change devices and enhance the granularity of the phase transitions, allowing for more intermediate levels. Each island will confine the phase transition to a smaller region, thus reducing the cumulative thermal and mechanical stress during switching. This localized switching can minimize the risk of material fatigue, defect accumulation, and degradation of the PCM or surrounding layers, which are common causes of failure in continuous PCM segments \cite{delaney2020new, teo2023capping, laprais2024reversible}. This approach may allow individual islands to be selectively switched, enabling partial operation and reducing the total number of switching events each island experiences, thereby prolonging the overall device lifetime. However, it may require more careful design and complex fabrication processes to ensure uniform switching behavior across all PCM islands. To achieve more operation levels, we can engineer the pulse shape to allow more precise control of the PCMs, such as multi-pulse techniques or pulse width modulation\cite{chen2023non}. Furthermore, increasing the capping layer thickness or employing other more inert capping materials such as silicon nitride can also increase the endurance of PCM\cite{martin2022endurance}.

It is important to note that the main novelty of this work lies in the demonstration of 127-level, low-energy hybrid tuning of micro-ring resonators, marrying  nonvolatile PCMs for coarse but energy efficient tuning and thermo-optic effect for precise tuning. It is well-known that PCM tuning is not deterministic for intermediate levels, due to cycle-to-cycle structural differences. Bringing the deterministic thermo-optic tuning to fine tune the response can overcome this long-standing limitation. On the other hand, compared to the pure thermo-optic approach, such hybrid approach has a much lower static power – the device is firstly brought closer to the desired operation point and then applied with only slight voltage for thermal fine tuning. This will easily provide orders of magnitude lower average static power, depending on the levels achievable with PCMs and system’s static time. We also note that since the applied electric current for thermo-optic tuning is only tiny, the loss from the free-carrier dispersion effect as well as the thermal crosstalk is significantly reduced. As another important achievement, we achieved operation voltage compatible with CMOS driving voltages (< 3V) by optimizing microheater geometry, allowing for integration with CMOS EICs using various integration techniques. This is a crucial step for the further scale up of PCM-based programmable PICs for applications in data centers and AI.

In summary, we have demonstrated low-voltage and low-energy switching of PCMs by carefully engineering the PIN doped silicon microheaters. We numerically and experimentally studied the heating performance of PIN diode heaters and observed a monotonically increasing switching voltage with the increase in intrinsic region width. With a small intrinsic region of 0.9 µm, we demonstrate CMOS-compatible switching voltage of 2.75V (1.6V), energy 35.33 nJ (0.48 mJ) for amorphization (crystallization) of a 10-µm-long \sbs on silicon waveguides. We further showed 127-level operation of PCM-clad Si micro-ring resonator and over 10 thousand cycles in total by combining PCM coarse tuning with thermo-optic fine tuning. Our work on low-voltage, low-energy, hybrid tuning of PCM photonic devices bridges the gap of CMOS compatible PCM switching, paving the way for scalable, co-packaged PCM-based electro-optical systems that meet the demands of AI, data centers, and neuromorphic computing.

\section{\label{sec:level1}METHODS}
\subsection{\label{sec:level2}SOI Device Fabrication}
The fabrication of silicon photonic devices was conducted on a commercial SOI Wafer featuring a 220 nm thick silicon layer atop a \SI{2}{\micro\metre} thick SiO2 layer (WaferPro). Electron beam lithography (EBL, JEOL JBX-6300FS) was used to write the device patterns using a positive-tone EBL resist (ZEP-520A), followed by partial etching to a depth of $\sim$ 120 nm using a fluorine-based inductively coupled plasma etcher (ICP, Oxford PlasmaLab 100 ICP-18) with a mix of SF6 and C4F8 gases, achieving an etch rate of $\sim$ 2.8 nm/sec. The doping regions were formed through two additional EBL processes with PMMA and removal of the surface native oxide layer was performed by immersing the chips in a 10:1 buffered oxide etchant (BOE) for 10 seconds, after which metal contacts were patterned through a fourth EBL step with PMMA. Metal deposition was achieved via electron-beam evaporation (CHA SEC-600) and subsequent lift-off of Ti/Pd (5 nm/180 nm). Measurements of PIN diode with the variation in intrinsic region from 0.9 to \SI{1.9}{\micro\metre}indicated a threshold voltage of approximately 0.8 V and a resistivity around 50 $\Omega$. After a fifth EBL step to define the ${Sb_2}S_3$ window, a 40 nm ${Sb_2}S_3$ thin film was deposited from a \sbs target (Plasmaterial) using a magnetron sputtering system (Lesker Lab 18), followed by a lift-off process. The ${Sb_2}S_3$ layer was then encapsulated with a 40 nm thick ${Al_2}O_3$ coating using thermal ALD (Oxford Plasmalab 80PLUS OpAL ALD) at 150°C. The detailed fabrication process can be referred from the cited paper\cite{chen2023non}. 
\subsection{\label{sec:level2}Heat Transfer Simulation}
The power requirement for reversible switching of the doped silicon PIN heater was simulated using COMSOL Multiphysics, a commercial simulation software \cite{zheng2020nonvolatile,zheng2020modeling,chen2023non}. This simulation integrated a heat transfer model in solids with a semiconductor model to evaluate the transient response of the device.
\subsection{\label{sec:level2}Optical transmission measurement setup}
The optical measurement setup consist of vertical fiber-coupling setup angled at 25 degrees, a thermoelectric controller (TEC, TE Technology TC-720), tunable continuous-wave laser (Santec TSL-510), manual fiber polarization controller (Thorlabs FPC526), a low-noise power meter (Keysight 81634B) to measure the static optical transmission. TEC was used to maintain the stage temperature at 26°C. Input light was provided by Santec TSL-510 with its polarization optimized by the polarization controller to achieve maximum fiber-to-chip coupling efficiency. For on-chip electrical switching, electrical pulses were applied to the on-chip metal contacts via a pair of electrical probes positioned with probe positioners (Cascade Microtech DPP105-M-AI-S). Crystallization and amorphization pulses were generated using a pulse function arbitrary generator (Keysight 81160 A). The tunable laser, power meter, thermal controller, source meter, and pulse function arbitrary generator were all controlled through a LabView program \cite{zheng2020nonvolatile}. The detailed optical measurement setup can be referred from this cited paper\cite{chen2023non}.

\begin{acknowledgments}
The research is funded by ONR-YIP Award, DARPA-YFA Award , Intel, and DARPA-NAPSACK. Part of this work was conducted at the Washington Nanofabrication Facility/ Molecular Analysis Facility, a National Nanotechnology Coordinated Infrastructure (NNCI) site at the University of Washington, with partial support from the National Science Foundation via awards NNCI-1542101 and NNCI-2025489. Part of this work was performed at the Stanford Nano Shared Facilities (SNSF), supported by the National Science Foundation under award ECCS-1542152).
\end{acknowledgments}

\section*{Data Availability Statement}

The data that support the findings of this study are available from the corresponding author upon request.



\section*{REFERENCES}
\nocite{*}
\bibliographystyle{unsrt} 
\bibliography{aipsamp}

\begin{thebibliography}{10}

\bibitem{zhu2023intelligent}
Shiqiang Zhu, Ting Yu, Tao Xu, Hongyang Chen, Schahram Dustdar, Sylvain Gigan, Deniz Gunduz, Ekram Hossain, Yaochu Jin, Feng Lin, et~al.
\newblock Intelligent computing: the latest advances, challenges, and future.
\newblock {\em Intelligent Computing}, 2:0006, 2023.

\bibitem{christensen20222022}
Dennis~V Christensen, Regina Dittmann, Bernabe Linares-Barranco, Abu Sebastian, Manuel Le~Gallo, Andrea Redaelli, Stefan Slesazeck, Thomas Mikolajick, Sabina Spiga, Stephan Menzel, et~al.
\newblock 2022 roadmap on neuromorphic computing and engineering.
\newblock {\em Neuromorphic Computing and Engineering}, 2(2):022501, 2022.

\bibitem{gill2022ai}
Sukhpal~Singh Gill, Minxian Xu, Carlo Ottaviani, Panos Patros, Rami Bahsoon, Arash Shaghaghi, Muhammed Golec, Vlado Stankovski, Huaming Wu, Ajith Abraham, et~al.
\newblock Ai for next generation computing: Emerging trends and future directions.
\newblock {\em Internet of Things}, 19:100514, 2022.

\bibitem{mayer2020scalable}
Ruben Mayer and Hans-Arno Jacobsen.
\newblock Scalable deep learning on distributed infrastructures: Challenges, techniques, and tools.
\newblock {\em ACM Computing Surveys (CSUR)}, 53(1):1--37, 2020.

\bibitem{kelechi2020artificial}
Anabi~Hilary Kelechi, Mohammed~H Alsharif, Okpe~Jonah Bameyi, Paul~Joan Ezra, Iorshase~Kator Joseph, Aaron-Anthony Atayero, Zong~Woo Geem, and Junhee Hong.
\newblock Artificial intelligence: An energy efficiency tool for enhanced high performance computing.
\newblock {\em Symmetry}, 12(6):1029, 2020.

\bibitem{markov2014limits}
Igor~L Markov.
\newblock Limits on fundamental limits to computation.
\newblock {\em Nature}, 512(7513):147--154, 2014.

\bibitem{thraskias2018survey}
Christos~A Thraskias, Eythimios~N Lallas, Niels Neumann, Laurent Schares, Bert~J Offrein, Ronny Henker, Dirk Plettemeier, Frank Ellinger, Juerg Leuthold, and Ioannis Tomkos.
\newblock Survey of photonic and plasmonic interconnect technologies for intra-datacenter and high-performance computing communications.
\newblock {\em IEEE Communications Surveys \& Tutorials}, 20(4):2758--2783, 2018.

\bibitem{tsiokos20162}
D~Tsiokos and GT~Kanellos.
\newblock 2.1 optical interconnects: the driver behind future data centers.
\newblock {\em Optical Interconnects for Data Centers}, page~43, 2016.

\bibitem{tsiokos2017optical}
D~Tsiokos and GT~Kanellos.
\newblock Optical interconnects: fundamentals.
\newblock In {\em Optical Interconnects for Data Centers}, pages 43--73. Elsevier, 2017.

\bibitem{haffner2015all}
Christian Haffner, Wolfgang Heni, Yuriy Fedoryshyn, Jens Niegemann, Argishti Melikyan, Delwin~L Elder, Benedikt Baeuerle, Yannick Salamin, Arne Josten, Ueli Koch, et~al.
\newblock All-plasmonic mach--zehnder modulator enabling optical high-speed communication at the microscale.
\newblock {\em Nature Photonics}, 9(8):525--528, 2015.

\bibitem{urata2022mission}
Ryohei Urata, Hong Liu, Kevin Yasumura, Erji Mao, Jill Berger, Xiang Zhou, Cedric Lam, Roy Bannon, Darren Hutchinson, Daniel Nelson, et~al.
\newblock Mission apollo: landing optical circuit switching at datacenter scale.
\newblock {\em arXiv preprint arXiv:2208.10041}, 2022.

\bibitem{liu2021bridging}
Zhengchun Liu, Ahsan Ali, Peter Kenesei, Antonino Miceli, Hemant Sharma, Nicholas Schwarz, Dennis Trujillo, Hyunseung Yoo, Ryan Coffee, Naoufal Layad, et~al.
\newblock Bridging data center ai systems with edge computing for actionable information retrieval.
\newblock In {\em 2021 3rd Annual Workshop on Extreme-scale Experiment-in-the-Loop Computing (XLOOP)}, pages 15--23. IEEE, 2021.

\bibitem{amirsoleimani2020memory}
Amirali Amirsoleimani, Fabien Alibart, Victor Yon, Jianxiong Xu, M~Reza Pazhouhandeh, Serge Ecoffey, Yann Beilliard, Roman Genov, and Dominique Drouin.
\newblock In-memory vector-matrix multiplication in monolithic complementary metal--oxide--semiconductor-memristor integrated circuits: Design choices, challenges, and perspectives.
\newblock {\em Advanced Intelligent Systems}, 2(11):2000115, 2020.

\bibitem{woods2017approximate}
Walt Woods and Christof Teuscher.
\newblock Approximate vector matrix multiplication implementations for neuromorphic applications using memristive crossbars.
\newblock In {\em 2017 IEEE/ACM International Symposium on Nanoscale Architectures (NANOARCH)}, pages 103--108. IEEE, 2017.

\bibitem{maslej2023artificial}
Nestor Maslej, Loredana Fattorini, Erik Brynjolfsson, John Etchemendy, Katrina Ligett, Terah Lyons, James Manyika, Helen Ngo, Juan~Carlos Niebles, Vanessa Parli, et~al.
\newblock Artificial intelligence index report 2023.
\newblock {\em arXiv preprint arXiv:2310.03715}, 2023.

\bibitem{zhou2022photonic}
Hailong Zhou, Jianji Dong, Junwei Cheng, Wenchan Dong, Chaoran Huang, Yichen Shen, Qiming Zhang, Min Gu, Chao Qian, Hongsheng Chen, et~al.
\newblock Photonic matrix multiplication lights up photonic accelerator and beyond.
\newblock {\em Light: Science \& Applications}, 11(1):30, 2022.

\bibitem{hattori2021neural}
Naoki Hattori, Jun Shiomi, Yutaka Masuda, Tohru Ishihara, Akihiko Shinya, and Masaya Notomi.
\newblock Neural network calculations at the speed of light using optical vector-matrix multiplication and optoelectronic activation.
\newblock {\em IEICE Transactions on Fundamentals of Electronics, Communications and Computer Sciences}, 104(11):1477--1487, 2021.

\bibitem{athale1982optical}
Ravindra~A Athale and William~C Collins.
\newblock Optical matrix--matrix multiplier based on outer product decomposition.
\newblock {\em Applied optics}, 21(12):2089--2090, 1982.

\bibitem{ocampo2024new}
Carlos A~R{\'\i}os Ocampo, Yifei Zhang, Mikhail Shalaginov, Tian Gu, and Juejun Hu.
\newblock New phase-change materials for photonic computing and beyond.
\newblock In {\em Phase Change Materials-Based Photonic Computing}, pages 145--192. Elsevier, 2024.

\bibitem{chen2024low}
Rui Chen, Virat Tara, Jayita Dutta, Zhuoran Fang, Jiajiu Zheng, and Arka Majumdar.
\newblock Low-loss multilevel operation using lossy phase-change material-integrated silicon photonics.
\newblock {\em Journal of Optical Microsystems}, 4(3):031202--031202, 2024.

\bibitem{zhu2020photonic}
Ziyi Zhu, Giuseppe Di~Guglielmo, Qixiang Cheng, Madeleine Glick, Jihye Kwon, Hang Guan, Luca~P Carloni, and Keren Bergman.
\newblock Photonic switched optically connected memory: an approach to address memory challenges in deep learning.
\newblock {\em Journal of Lightwave Technology}, 38(10):2815--2825, 2020.

\bibitem{chakraborty2019photonic}
Indranil Chakraborty, Gobinda Saha, and Kaushik Roy.
\newblock Photonic in-memory computing primitive for spiking neural networks using phase-change materials.
\newblock {\em Physical Review Applied}, 11(1):014063, 2019.

\bibitem{lian2022photonic}
Chuanyu Lian, Christos Vagionas, Theonitsa Alexoudi, Nikos Pleros, Nathan Youngblood, and Carlos R{\'\i}os.
\newblock Photonic (computational) memories: tunable nanophotonics for data storage and computing.
\newblock {\em Nanophotonics}, 11(17):3823--3854, 2022.

\bibitem{shen2017deep}
Yichen Shen, Nicholas~C Harris, Scott Skirlo, Mihika Prabhu, Tom Baehr-Jones, Michael Hochberg, Xin Sun, Shijie Zhao, Hugo Larochelle, Dirk Englund, et~al.
\newblock Deep learning with coherent nanophotonic circuits.
\newblock {\em Nature photonics}, 11(7):441--446, 2017.

\bibitem{capmany2016programmable}
Jos{\'e} Capmany, Ivana Gasulla, and Daniel P{\'e}rez.
\newblock The programmable processor.
\newblock {\em Nature Photonics}, 10(1):6--8, 2016.

\bibitem{harris2017quantum}
Nicholas~C Harris, Gregory~R Steinbrecher, Mihika Prabhu, Yoav Lahini, Jacob Mower, Darius Bunandar, Changchen Chen, Franco~NC Wong, Tom Baehr-Jones, Michael Hochberg, et~al.
\newblock Quantum transport simulations in a programmable nanophotonic processor.
\newblock {\em Nature Photonics}, 11(7):447--452, 2017.

\bibitem{zhu2021integrated}
Di~Zhu, Linbo Shao, Mengjie Yu, Rebecca Cheng, Boris Desiatov, CJ~Xin, Yaowen Hu, Jeffrey Holzgrafe, Soumya Ghosh, Amirhassan Shams-Ansari, et~al.
\newblock Integrated photonics on thin-film lithium niobate.
\newblock {\em Advances in Optics and Photonics}, 13(2):242--352, 2021.

\bibitem{prabhathan2023roadmap}
Patinharekandy Prabhathan, Kandammathe~Valiyaveedu Sreekanth, Jinghua Teng, Joo~Hwan Ko, Young~Jin Yoo, Hyeon-Ho Jeong, Yubin Lee, Shoujun Zhang, Tun Cao, Cosmin-Constantin Popescu, et~al.
\newblock Roadmap for phase change materials in photonics and beyond.
\newblock {\em Iscience}, 26(10), 2023.

\bibitem{moody20222022}
Galan Moody, Volker~J Sorger, Daniel~J Blumenthal, Paul~W Juodawlkis, William Loh, Cheryl Sorace-Agaskar, Alex~E Jones, Krishna~C Balram, Jonathan~CF Matthews, Anthony Laing, et~al.
\newblock 2022 roadmap on integrated quantum photonics.
\newblock {\em Journal of Physics: Photonics}, 4(1):012501, 2022.

\bibitem{schulz2024roadmap}
Sebastian~A Schulz, Rupert Oulton, Mitchell Kenney, Andrea Al{\`u}, Isabelle Staude, Ayesheh Bashiri, Zlata Fedorova, Radoslaw Kolkowski, A~Femius Koenderink, Xiaofei Xiao, et~al.
\newblock Roadmap on photonic metasurfaces.
\newblock {\em Applied Physics Letters}, 124(26), 2024.

\bibitem{bogaerts2020programmable}
Wim Bogaerts, Daniel P{\'e}rez, Jos{\'e} Capmany, David~AB Miller, Joyce Poon, Dirk Englund, Francesco Morichetti, and Andrea Melloni.
\newblock Programmable photonic circuits.
\newblock {\em Nature}, 586(7828):207--216, 2020.

\bibitem{haselman2009future}
Michael Haselman and Scott Hauck.
\newblock The future of integrated circuits: A survey of nanoelectronics.
\newblock {\em Proceedings of the IEEE}, 98(1):11--38, 2009.

\bibitem{geler2022ferroelectric}
Jacqueline Geler-Kremer, Felix Eltes, Pascal Stark, David Stark, Daniele Caimi, Heinz Siegwart, Bert Jan~Offrein, Jean Fompeyrine, and Stefan Abel.
\newblock A ferroelectric multilevel non-volatile photonic phase shifter.
\newblock {\em Nature Photonics}, 16(7):491--497, 2022.

\bibitem{gao2021intermediate}
Kun Gao, Kang Du, Senmao Tian, Heng Wang, Lu~Zhang, Yaxuan Guo, Bingcheng Luo, Wending Zhang, and Ting Mei.
\newblock Intermediate phase-change states with improved cycling durability of sb2s3 by femtosecond multi-pulse laser irradiation.
\newblock {\em Advanced Functional Materials}, 31(35):2103327, 2021.

\bibitem{moitra2023programmable}
Parikshit Moitra, Yunzheng Wang, Xinan Liang, Li~Lu, Alyssa Poh, Tobias~WW Mass, Robert~E Simpson, Arseniy~I Kuznetsov, and Ramon Paniagua-Dominguez.
\newblock Programmable wavefront control in the visible spectrum using low-loss chalcogenide phase-change metasurfaces.
\newblock {\em Advanced Materials}, 35(34):2205367, 2023.

\bibitem{li2019fast}
Xuan Li, Nathan Youngblood, Carlos R{\'\i}os, Zengguang Cheng, C~David Wright, Wolfram~HP Pernice, and Harish Bhaskaran.
\newblock Fast and reliable storage using a 5 bit, nonvolatile photonic memory cell.
\newblock {\em Optica}, 6(1):1--6, 2019.

\bibitem{perez2017multipurpose}
D~P{\'e}rez, I~Gasulla, C~Lee, DJ~Thomson, AZ~Khokhar, K~Li, W~Cao, GZ~Mashanovich, and J~Capmany.
\newblock Multipurpose silicon photonics signal processor core nat, 2023.

\bibitem{zhang2020photonic}
Weifeng Zhang and Jianping Yao.
\newblock Photonic integrated field-programmable disk array signal processor.
\newblock {\em Nature communications}, 11(1):406, 2020.

\bibitem{reed2010silicon}
Graham~T Reed, Goran Mashanovich, F~Yand Gardes, and DJhttps Thomson.
\newblock Silicon optical modulators.
\newblock {\em Nature photonics}, 4(8):518--526, 2010.

\bibitem{wang2018integrated}
Cheng Wang, Mian Zhang, Xi~Chen, Maxime Bertrand, Amirhassan Shams-Ansari, Sethumadhavan Chandrasekhar, Peter Winzer, and Marko Lon{\v{c}}ar.
\newblock Integrated lithium niobate electro-optic modulators operating at cmos-compatible voltages.
\newblock {\em Nature}, 562(7725):101--104, 2018.

\bibitem{melikyan2014high}
Argishti Melikyan, Luca Alloatti, Alban Muslija, David Hillerkuss, Philipp~C Schindler, J~Li, Robert Palmer, Dietmar Korn, Sascha Muehlbrandt, Dries Van~Thourhout, et~al.
\newblock High-speed plasmonic phase modulators.
\newblock {\em Nature Photonics}, 8(3):229--233, 2014.

\bibitem{prencipe2023electro}
Alessandro Prencipe and Katia Gallo.
\newblock Electro-and thermo-optics response of x-cut thin film linbo 3 waveguides.
\newblock {\em IEEE Journal of Quantum Electronics}, 59(3):1--8, 2023.

\bibitem{schrauwen2008trimming}
Jonathan Schrauwen, Dries Van~Thourhout, and Roel Baets.
\newblock Trimming of silicon ring resonator by electron beam induced compaction and strain.
\newblock {\em Optics express}, 16(6):3738--3743, 2008.

\bibitem{feldmann2021parallel}
Johannes Feldmann, Nathan Youngblood, Maxim Karpov, Helge Gehring, Xuan Li, Maik Stappers, Manuel Le~Gallo, Xin Fu, Anton Lukashchuk, Arslan~S Raja, et~al.
\newblock Parallel convolutional processing using an integrated photonic tensor core.
\newblock {\em Nature}, 589(7840):52--58, 2021.

\bibitem{wu2021programmable}
Changming Wu, Heshan Yu, Seokhyeong Lee, Ruoming Peng, Ichiro Takeuchi, and Mo~Li.
\newblock Programmable phase-change metasurfaces on waveguides for multimode photonic convolutional neural network.
\newblock {\em Nature communications}, 12(1):96, 2021.

\bibitem{wan2020large}
Noel~H Wan, Tsung-Ju Lu, Kevin~C Chen, Michael~P Walsh, Matthew~E Trusheim, Lorenzo De~Santis, Eric~A Bersin, Isaac~B Harris, Sara~L Mouradian, Ian~R Christen, et~al.
\newblock Large-scale integration of artificial atoms in hybrid photonic circuits.
\newblock {\em Nature}, 583(7815):226--231, 2020.

\bibitem{zhang2021myths}
Yifei Zhang, Carlos R{\'\i}os, Mikhail~Y Shalaginov, Mo~Li, Arka Majumdar, Tian Gu, and Juejun Hu.
\newblock Myths and truths about optical phase change materials: A perspective.
\newblock {\em Applied Physics Letters}, 118(21), 2021.

\bibitem{wuttig2017phase}
Matthias Wuttig, Harish Bhaskaran, and Thomas Taubner.
\newblock Phase-change materials for non-volatile photonic applications.
\newblock {\em Nature photonics}, 11(8):465--476, 2017.

\bibitem{abdollahramezani2020tunable}
Sajjad Abdollahramezani, Omid Hemmatyar, Hossein Taghinejad, Alex Krasnok, Yashar Kiarashinejad, Mohammadreza Zandehshahvar, Andrea Al{\`u}, and Ali Adibi.
\newblock Tunable nanophotonics enabled by chalcogenide phase-change materials.
\newblock {\em Nanophotonics}, 9(5):1189--1241, 2020.

\bibitem{fang2021non}
Zhuoran Fang, Rui Chen, Jiajiu Zheng, and Arka Majumdar.
\newblock Non-volatile reconfigurable silicon photonics based on phase-change materials.
\newblock {\em IEEE Journal of Selected Topics in Quantum Electronics}, 28(3: Hybrid Integration for Silicon Photonics):1--17, 2021.

\bibitem{chen2023non}
Rui Chen, Zhuoran Fang, Christopher Perez, Forrest Miller, Khushboo Kumari, Abhi Saxena, Jiajiu Zheng, Sarah~J Geiger, Kenneth~E Goodson, and Arka Majumdar.
\newblock Non-volatile electrically programmable integrated photonics with a 5-bit operation.
\newblock {\em Nature Communications}, 14(1):3465, 2023.

\bibitem{chen2022opportunities}
Rui Chen, Zhuoran Fang, Forrest Miller, Hannah Rarick, Johannes~E Froch, and Arka Majumdar.
\newblock Opportunities and challenges for large-scale phase-change material integrated electro-photonics.
\newblock {\em ACS Photonics}, 9(10):3181--3195, 2022.

\bibitem{rios2022ultra}
Carlos R{\'\i}os, Qingyang Du, Yifei Zhang, Cosmin-Constantin Popescu, Mikhail~Y Shalaginov, Paul Miller, Christopher Roberts, Myungkoo Kang, Kathleen~A Richardson, Tian Gu, et~al.
\newblock Ultra-compact nonvolatile phase shifter based on electrically reprogrammable transparent phase change materials.
\newblock {\em PhotoniX}, 3(1):26, 2022.

\bibitem{zheng2018gst}
Jiajiu Zheng, Amey Khanolkar, Peipeng Xu, Shane Colburn, Sanchit Deshmukh, Jason Myers, Jesse Frantz, Eric Pop, Joshua Hendrickson, Jonathan Doylend, et~al.
\newblock Gst-on-silicon hybrid nanophotonic integrated circuits: a non-volatile quasi-continuously reprogrammable platform.
\newblock {\em Optical Materials Express}, 8(6):1551--1561, 2018.

\bibitem{xu2019low}
Peipeng Xu, Jiajiu Zheng, Jonathan~K Doylend, and Arka Majumdar.
\newblock Low-loss and broadband nonvolatile phase-change directional coupler switches.
\newblock {\em Acs Photonics}, 6(2):553--557, 2019.

\bibitem{chen2022broadband}
Rui Chen, Zhuoran Fang, Johannes~E Froch, Peipeng Xu, Jiajiu Zheng, and Arka Majumdar.
\newblock Broadband nonvolatile electrically controlled programmable units in silicon photonics.
\newblock {\em ACS Photonics}, 9(6):2142--2150, 2022.

\bibitem{fang2022ultra}
Zhuoran Fang, Rui Chen, Jiajiu Zheng, Asir~Intisar Khan, Kathryn~M Neilson, Sarah~J Geiger, Dennis~M Callahan, Michael~G Moebius, Abhi Saxena, Michelle~E Chen, et~al.
\newblock Ultra-low-energy programmable non-volatile silicon photonics based on phase-change materials with graphene heaters.
\newblock {\em Nature Nanotechnology}, 17(8):842--848, 2022.

\bibitem{zheng2020nonvolatile}
Jiajiu Zheng, Zhuoran Fang, Changming Wu, Shifeng Zhu, Peipeng Xu, Jonathan~K Doylend, Sanchit Deshmukh, Eric Pop, Scott Dunham, Mo~Li, et~al.
\newblock Nonvolatile electrically reconfigurable integrated photonic switch enabled by a silicon pin diode heater.
\newblock {\em Advanced Materials}, 32(31):2001218, 2020.

\bibitem{rios2015integrated}
Carlos R{\'\i}os, Matthias Stegmaier, Peiman Hosseini, Di~Wang, Torsten Scherer, C~David Wright, Harish Bhaskaran, and Wolfram~HP Pernice.
\newblock Integrated all-photonic non-volatile multi-level memory.
\newblock {\em Nature photonics}, 9(11):725--732, 2015.

\bibitem{erickson2022designing}
John~R Erickson, Vivswan Shah, Qingzhou Wan, Nathan Youngblood, and Feng Xiong.
\newblock Designing fast and efficient electrically driven phase change photonics using foundry compatible waveguide-integrated microheaters.
\newblock {\em Optics Express}, 30(8):13673--13689, 2022.

\bibitem{chen2024deterministic}
Rui Chen, Virat Tara, Minho Choi, Jayita Dutta, Justin Sim, Julian Ye, Zhuoran Fang, Jiajiu Zheng, and Arka Majumdar.
\newblock Deterministic quasi-continuous tuning of phase-change material integrated on a high-volume 300-mm silicon photonics platform.
\newblock {\em npj Nanophotonics}, 1(1):7, 2024.

\bibitem{chen2024toward}
Rui Chen, Virat Tara, Jayita Duta, Minho Choi, Justin Sim, Julian Ye, Jiajiu Zheng, Zhuoran Fang, and Arka Majumdar.
\newblock Toward large-scale nonvolatile electrical programmable photonics with deterministic multilevel operation.
\newblock In {\em Optical Fiber Communication Conference}, pages M4A--4. Optica Publishing Group, 2024.

\bibitem{zhang2019broadband}
Yifei Zhang, Jeffrey~B Chou, Junying Li, Huashan Li, Qingyang Du, Anupama Yadav, Si~Zhou, Mikhail~Y Shalaginov, Zhuoran Fang, Huikai Zhong, et~al.
\newblock Broadband transparent optical phase change materials for high-performance nonvolatile photonics.
\newblock {\em Nature communications}, 10(1):4279, 2019.

\bibitem{delaney2020new}
Matthew Delaney, Ioannis Zeimpekis, Daniel Lawson, Daniel~W Hewak, and Otto~L Muskens.
\newblock A new family of ultralow loss reversible phase-change materials for photonic integrated circuits: Sb2s3 and sb2se3.
\newblock {\em Advanced functional materials}, 30(36):2002447, 2020.

\bibitem{vcernovskova2014thermal}
E~{\v{C}}erno{\v{s}}kov{\'a}, R~Todorov, Z~{\v{C}}erno{\v{s}}ek, J~Holubov{\'a}, and L~Bene{\v{s}}.
\newblock Thermal properties and the structure of amorphous sb 2 se 3 thin film.
\newblock {\em Journal of thermal analysis and calorimetry}, 118:105--110, 2014.

\bibitem{dong2019wide}
Weiling Dong, Hailong Liu, Jitendra~K Behera, Li~Lu, Ray~JH Ng, Kandammathe~Valiyaveedu Sreekanth, Xilin Zhou, Joel~KW Yang, and Robert~E Simpson.
\newblock Wide bandgap phase change material tuned visible photonics.
\newblock {\em Advanced Functional Materials}, 29(6):1806181, 2019.

\bibitem{nasr2016first}
T~Ben Nasr, H~Maghraoui-Meherzi, and N~Kamoun-Turki.
\newblock First-principles study of electronic, thermoelectric and thermal properties of sb2s3.
\newblock {\em Journal of Alloys and Compounds}, 663:123--127, 2016.

\bibitem{gutierrez2022interlaboratory}
Yael Guti{\'e}rrez, Anna~P Ovvyan, Gonzalo Santos, Dilson Juan, Saul~A Rosales, Javier Junquera, Pablo Garc{\'\i}a-Fern{\'a}ndez, Stefano Dicorato, Maria~M Giangregorio, Elena Dilonardo, et~al.
\newblock Interlaboratory study on sb2s3 interplay between structure, dielectric function, and amorphous-to-crystalline phase change for photonics.
\newblock {\em Iscience}, 25(6), 2022.

\bibitem{wei2023electrically}
Maoliang Wei, Junying Li, Zequn Chen, Bo~Tang, Zhiqi Jia, Peng Zhang, Kunhao Lei, Kai Xu, Jianghong Wu, Chuyu Zhong, et~al.
\newblock Electrically programmable phase-change photonic memory for optical neural networks with nanoseconds in situ training capability.
\newblock {\em Advanced Photonics}, 5(4):046004, 2023.

\bibitem{shalaginov2021reconfigurable}
Mikhail~Y Shalaginov, Sensong An, Yifei Zhang, Fan Yang, Peter Su, Vladimir Liberman, Jeffrey~B Chou, Christopher~M Roberts, Myungkoo Kang, Carlos Rios, et~al.
\newblock Reconfigurable all-dielectric metalens with diffraction-limited performance.
\newblock {\em Nature communications}, 12(1):1225, 2021.

\bibitem{lu2021reversible}
Li~Lu, Zhaogang Dong, Febiana Tijiptoharsono, Ray Jia~Hong Ng, Hongtao Wang, Soroosh~Daqiqeh Rezaei, Yunzheng Wang, Hai~Sheng Leong, Poh~Chong Lim, Joel~KW Yang, et~al.
\newblock Reversible tuning of mie resonances in the visible spectrum.
\newblock {\em ACS nano}, 15(12):19722--19732, 2021.

\bibitem{zhang2018broadband}
Qihang Zhang, Yifei Zhang, Junying Li, Richard Soref, Tian Gu, and Juejun Hu.
\newblock Broadband nonvolatile photonic switching based on optical phase change materials: beyond the classical figure-of-merit.
\newblock {\em Optics letters}, 43(1):94--97, 2018.

\bibitem{spallholz1994nature}
Julian~E Spallholz.
\newblock On the nature of selenium toxicity and carcinostatic activity.
\newblock {\em Free Radical Biology and Medicine}, 17(1):45--64, 1994.

\bibitem{zhang2023nonvolatile}
Changping Zhang, Maoliang Wei, Jun Zheng, Shujun Liu, Hongyuan Cao, Yishu Huang, Ying Tan, Ming Zhang, Yiwei Xie, Zejie Yu, et~al.
\newblock Nonvolatile multilevel switching of silicon photonic devices with in2o3/gst segmented structures.
\newblock {\em Advanced Optical Materials}, 11(8):2202748, 2023.

\bibitem{tuma2016stochastic}
Tomas Tuma, Angeliki Pantazi, Manuel Le~Gallo, Abu Sebastian, and Evangelos Eleftheriou.
\newblock Stochastic phase-change neurons.
\newblock {\em Nature nanotechnology}, 11(8):693--699, 2016.

\bibitem{fang2023non}
Zhuoran Fang, Rui Chen, Bassem Tossoun, Stanley Cheung, Di~Liang, and Arka Majumdar.
\newblock Non-volatile materials for programmable photonics.
\newblock {\em APL Materials}, 11(10), 2023.

\bibitem{laprais2024reversible}
Capucine Laprais, Cl{\'e}ment Zrounba, Julien Bouvier, Nicholas Blanchard, Matthieu Bugnet, Alban Gassenq, Yael Guti{\'e}rrez, Saul Vazquez-Miranda, Shirly Espinoza, Peter Thiesen, et~al.
\newblock Reversible single-pulse laser-induced phase change of sb2s3 thin films: Multi-physics modeling and experimental demonstrations.
\newblock {\em Advanced Optical Materials}, page 2401214, 2024.

\bibitem{aryana2023toward}
Kiumars Aryana, Hyun~Jung Kim, Cosmin-Constantin Popescu, Steven Vitale, Hyung~Bin Bae, Taewoo Lee, Tian Gu, and Juejun Hu.
\newblock Toward accurate thermal modeling of phase change material-based photonic devices.
\newblock {\em Small}, 19(50):2304145, 2023.

\bibitem{bal2021nucleation}
Kristof~M Bal.
\newblock Nucleation rates from small scale atomistic simulations and transition state theory.
\newblock {\em The Journal of Chemical Physics}, 155(14), 2021.

\bibitem{han2023energy}
Shangtong Han, Ruiqing He, Changping Zhang, Huan Li, Daoxin Dai, and Yaocheng Shi.
\newblock Energy-efficient nonvolatile switching of silicon microring resonator with suspended phase-change waveguide.
\newblock In {\em 2023 Opto-Electronics and Communications Conference (OECC)}, pages 1--2. IEEE, 2023.

\bibitem{teo2023capping}
Ting~Yu Teo, Nanxi Li, Landobasa~YM Tobing, Amy Sen~Kay Tong, Doris Keh~Ting Ng, Zhihao Ren, Chengkuo Lee, Lennon Yao~Ting Lee, and Robert~E Simpson.
\newblock Capping layer effects on sb2s3-based reconfigurable photonic devices.
\newblock {\em ACS Photonics}, 10(9):3203--3214, 2023.

\bibitem{martin2022endurance}
Louis Martin-Monier, Cosmin~Constantin Popescu, Luigi Ranno, Brian Mills, Sarah Geiger, Dennis Callahan, Michael Moebius, and Juejun Hu.
\newblock Endurance of chalcogenide optical phase change materials: a review.
\newblock {\em Optical Materials Express}, 12(6):2145--2167, 2022.

\bibitem{zheng2020modeling}
Jiajiu Zheng, Shifeng Zhu, Peipeng Xu, Scott Dunham, and Arka Majumdar.
\newblock Modeling electrical switching of nonvolatile phase-change integrated nanophotonic structures with graphene heaters.
\newblock {\em ACS applied materials \& interfaces}, 12(19):21827--21836, 2020.

\bibitem{novarese2022study}
Marco Novarese, Sebastian~Romero Garcia, Stefania Cucco, Don Adams, Jock Bovington, and Mariangela Gioannini.
\newblock Study of nonlinear effects and self-heating in a silicon microring resonator including a shockley-read-hall model for carrier recombination.
\newblock {\em Optics Express}, 30(9):14341--14357, 2022.

\bibitem{bao2023harnessing}
Peng Bao, Qixiang Cheng, Jinlong Wei, Giuseppe Talli, Maxim Kuschnerov, and Richard~V Penty.
\newblock Harnessing self-heating effect for ultralow-crosstalk electro-optic mach--zehnder switches.
\newblock {\em Photonics Research}, 11(10):1757--1769, 2023.

\end{thebibliography}

\end{document}